\documentclass[12pt]{article}
\usepackage{fleqn}
\usepackage{graphicx} 

\begin{document}

\title{{\bf Nuclear disintegration induced by virtual photons at heavy-ion
colliders}}
    
\author{I.A. Pshenichnov
\thanks{Supported by INTAS fellowship YSF-98-86}
\thanks{e-mail: PSHENICHNOV@AL20.INR.TROITSK.RU}\\
{\em Institute for Nuclear Research, Russian Academy of Science} \\
{\em 117312 Moscow, Russia}
}

\date{}

\maketitle

\begin{abstract}
A model of electromagnetic interaction of relativistic heavy ions which
makes it possible to calculate the inclusive and exclusive 
characteristics of such interactions is presented. The multistep model 
includes (1) the single and double virtual photon absorption by a nucleus,
(2) intranuclear cascades of produced hadrons and (3) statistical
decay of excited residual nuclei described by evaporation, fission and 
multifragmentation mechanisms. Components of the model were verified by 
comparison of calculation results with experimental data obtained with
monoenergetic photons and heavy ions at KEK-Tanashi Synchrotron and 
CERN SPS accelerator, respectively.   
\end{abstract}

\section{Description of the model}
The relativistic Coulomb excitation of nuclei is a well-known 
phenomenon~\cite{BertBaur}.  It is expected to play an important role
in collisions of ultrarelativistic heavy nuclei at new colliders, 
RHIC and LHC~\cite{BaurHen,Krauss}. The process may be described in the 
framework of the Weizs\"{a}cker-Williams method of equivalent photons. 
According to this method the impact of Lorentz-boosted Coulomb field 
of a nucleus on the collision partner may be treated as the absorption of
equivalent photons. Due to the coherent nature of photon 
emission by the nucleus, such photons are almost real, 
$Q^2\leq 1/R^2$, $R$ is the
nuclear radius. Therefore photonuclear reaction data obtained in experiments
with real monoenergetic photons and appropriate theoretical
models describing photonuclear reactions may be used to predict the 
properties of electromagnetic dissociation process. 
 
Since an ultrarelativistic nucleus moving at the impact parameter  $b$ 
with the velocity $\beta=v/c\approx 1$, ($\gamma\gg 1$)  
spends a short time  $\Delta t$  near the 
collision partner, the virtual photon spectrum contains all the frequencies up 
to the maximum energy $E_{\gamma}^{max}\sim 1/\Delta t\sim 
\gamma/R$:
\begin{equation}
N(E_\gamma ,b)=\frac{\alpha Z^{2}_t}{\pi ^2}
\frac{{\sf x}^2}{\beta ^2 E_\gamma b^2} 
\Bigl(K^{2}_{1}({\sf x})+\frac{1}{\gamma ^2}K^{2}_{0}({\sf x})\Bigl),
\end{equation}
\noindent where $\alpha$ is the fine structure constant,  
$K_0$ and $K_1$ are the modified Bessel functions and 
${\sf x}=E_\gamma b/(\gamma \beta)$. The mean number of photons absorbed 
by the collision partner of mass $A_p$  is defined by:
\begin{equation}
m(b)=
\int\limits_{E_{\gamma}^{min}}^{E_{\gamma}^{max}}N(E_\gamma ,b)
\sigma_{A_p}(E_\gamma)dE_\gamma ,
\label{eq:2}
\end{equation}
where $\sigma_{A_p}(E_\gamma)$ is the appropriate total photoabsorption 
cross section obtained  starting from the photoneutron threshold at 
$E_{\gamma}^{min}\sim 7$ MeV.

Assuming the Poison distribution for the multiphoton absorption 
with the mean multiplicity 
$m(b)$,~Eq.(\ref{eq:2}),
the integral cross 
sections  are calculated for the first- and second-order 
processes~\cite{Pshenichnov2} for a particular dissociation channel~$i$:
\begin{equation}
\sigma^{(1)}_{i}=
\int\limits_{E_{\gamma}^{min}}^{E_{\gamma}^{max}} dE_1 N^{(1)}(E_1)
\sigma_{A_p}(E_1)f^{(1)}_{i}(E_1),
\end{equation}
\begin{equation}
\sigma^{(2)}_{i}=
\int\limits_{E_{\gamma}^{min}}^{E_{\gamma}^{max}}\int\limits_{E_{\gamma}^{min}}
^{E_{\gamma}^{max}}
dE_1dE_2 N^{(2)}(E_1,E_2)\sigma_{A_p}(E_1)\sigma_{A_p}(E_2)
f^{(2)}_{i}(E_1,E_2),
\end{equation}
with the following single and double photon spectral functions:
\begin{eqnarray}
N^{(1)}(E_1)=2\pi\int\limits_{b_{min}}^{\infty} bdb e^{-m(b)} 
N(E_1,b),\nonumber\\ 
N^{(2)}(E_1,E_2)=\pi\int\limits_{b_{min}}^{\infty} bdb 
e^{-m(b)} N(E_1,b) N(E_2,b).\nonumber  
\end{eqnarray}
\noindent The values $f^{(1)}_{i}(E_1)$ and $f^{(2)}_{i}(E_1,E_2)$ defined 
as the branching ratios for the considered disintegration channel~$i$
in the single and double photon absorption, respectively, have to be calculated
by our multistep model. Similar expressions may be written for the differential
distributions of produced particles on the rapidity, transverse momentum
and other variables.

In order to obtain $f^{(1)}_{i}$ and $f^{(2)}_{i}$
several mechanisms are included in a new 
Relativistic ELectromagnetic DISsociation (RELDIS) code for the Monte Carlo
simulation, namely, the intranuclear cascade of fast particles produced 
after the photon absorption on a nucleon or nuclear pair~\cite{Iljinov}, 
and the evaporation of nucleons and lightest fragments, 
binary fission or multifragmentation~\cite{JPB} at a later stage 
of interaction. The multifragmentation process dominates when the
excitation energy of residual nucleus, $E^\star$, exceeds $3-4$~MeV/nucleon 
and is described by the statistical multifragmentation model (SMM)~\cite{JPB}. 
For fissile nuclei the evaporation may take place before or after the fission.
The competition of evaporation and fission is also described with the
SMM. The lightest fragments may be also created via 
coalescence of fast nucleons into d, t, $^3{\rm He}$ or 
$^4{\rm He}$~\cite{Sudov}.
The decay of highly excited light residual nuclei with $A\leq 16$ is 
treated by 
the Fermi break-up mechanism~\cite{JPB}. Other details of the calculation 
scheme may be found in Refs.~\cite{Pshenichnov2,Iljinov,Pshenichnov}.

\section{Emission of nucleons and lightest fragments}
Depending on the photon energy, $E_\gamma$, and  mass number, $A_p$, 
different processes take place in the nuclear photoabsorption. 
Due to the contributions of several mechanisms: $\gamma N\rightarrow\pi N\/$,
$\gamma N\rightarrow2\pi N\/$ and $\gamma (np)\rightarrow n p\ $
the calculated double differential spectra 
$d^2\sigma /d\Omega dP$ of pions and 
protons have complex shapes above the pion production 
threshold~\cite{Iljinov}. The spectra of fast particles $\pi^+,\pi^-,\eta$ and
$p$ predicted by the model were compared in Ref.~\cite{Iljinov} 
with available sets of experimental data at $140\leq E_\gamma\leq 1000$ MeV 
including the data obtained with KEK-Tanashi, the 1.3-GeV Electron 
Synchrotron. A satisfactory description of the spectra was obtained. 
  
Fast hadrons produced after the photon absorption initiate a cascade
of subsequent collisions with the intranuclear nucleons leading to the
heating of a residual nucleus. Later the nucleus undergoes de-excitation by
means of the emission of nucleons and fragments. Because of a low Coulomb
barrier in light nuclei the rates of proton and neutron emission are 
comparable. 
  
\begin{figure}[htb]
\begin{center}
{\includegraphics[width=\textwidth]
{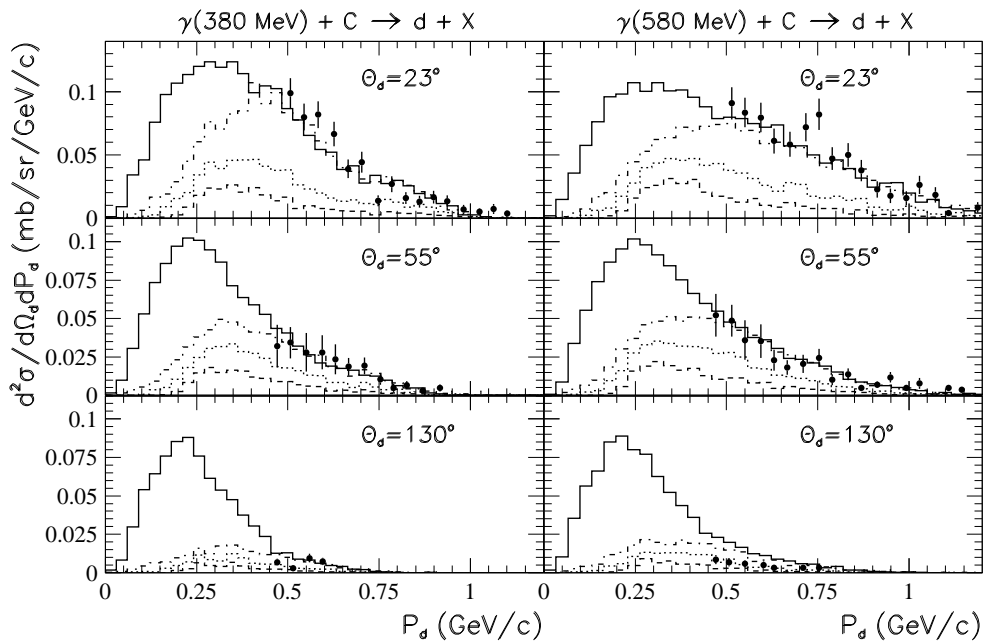}}
\end{center}
\caption{Deuteron emission in 
photoabsorption on carbon.  Points: KEK-Tanashi  data~\protect\cite{Baba}.
Solid histograms: results of the intranuclear cascade calculations
with coalescence and Fermi break-up. Contributions
from coalescence mechanism are shown by the dashed, dotted and dash-dotted
histograms for different values of coalescence 
parameter~\protect\cite{Sudov}: $p_0 =90, 129$ and
200 MeV/c, respectively.}
\label{fig:C}
\end{figure}

Fragment spectra in the photoabsorption on a carbon nucleus are given in 
Fig.~\ref{fig:C}. The low energy part of the deuteron spectra
is explained by the explosive Fermi break-up while the high energy part 
is attributed to
the coalescence mechanism. Since the main part of fast nucleons is emitted in 
the forward direction, the coalescence contribution dominates at small angles.
On the contrary, the distribution of deuterons from Fermi break-up is
nearly isotropic.  
Because of the option to accelerate also light oxygen ions
at RHIC, the calculations of the electromagnetic dissociation   
contribution for such ions should take into account  
both of the considered mechanisms.   

The photoabsorption scenario is very different for heavy nuclei, gold 
and lead. In this case neutrons are the most abundant particles 
produced in the electromagnetic collisions of ultrarelativistic 
nuclei~\cite{Pshenichnov2}.
When a heavy nucleus absorbs photons in the 
Giant Resonance region, $6\leq E_\gamma \leq 30$ MeV, the evaporation model 
may be used with the assumption $E^\star=E_\gamma$ which leads to 
the emission of one or two neutrons. 

On the other side of the equivalent photon
spectrum, when $E_\gamma$ reaches the value of several GeV, the multiple
pion photoproduction on intranuclear nucleons becomes the main absorption 
mechanism. 
In this case up to 95\% 
of the photon energy  is released in the fast particles on average.
Nevertheless, the remaining energy deposited in the residual nucleus
is sufficient for evaporating many neutrons or even 
multifragmentation~\cite{Pshenichnov}. 
As it was found in Ref.~\cite{Pshenichnov2} for PbPb collisions at
SPS ($158A$ GeV beam) and  LHC ($2.75A+2.75A$ TeV beams)
the mean neutron multiplicities are 4.2 and 8.8, respectively.
The same value for AuAu collisions at RHIC ($100A+100A$ GeV) is equal to 7.2.

\begin{figure}[htb]
\begin{centering}
{\includegraphics[width=0.6\textwidth]{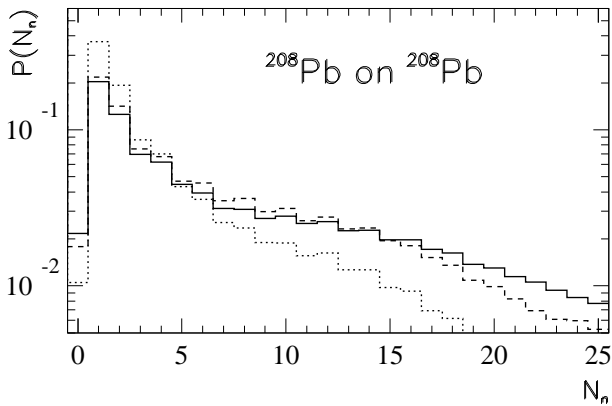}}
\caption{RELDIS predictions for multiplicity distributions of 
neutrons in the electromagnetic dissociation 
of Pb nuclei at LHC and SPS energies (solid and dotted lines,
respectively) and Au nuclei at RHIC energies (dashed lines).}
\end{centering}
\label{fig:nmult}
\end{figure}

The neutron multiplicity distributions are shown in 
Fig.~\ref{fig:nmult}. They are strongly peaked at the single neutron emission 
channel due to GDR decay, while there is a long tail of the multiple neutron 
emission originating from high excitations.
These results provide important information for designing large-rapidity 
detectors and zero-degree calorimeters at RHIC and LHC.  

\section{Charge changing reactions at SPS}

CERN SPS accelerator currently supplies the highest energy available for 
heavy ion studies. A detailed understanding of fragmentation mechanisms at SPS
energies provides an important point for the extrapolation of the theoretical 
and experimental results to RHIC and LHC energies. 
\begin{figure}[htb]
\begin{minipage}[t]{69mm}
{\includegraphics[width=0.9\textwidth]{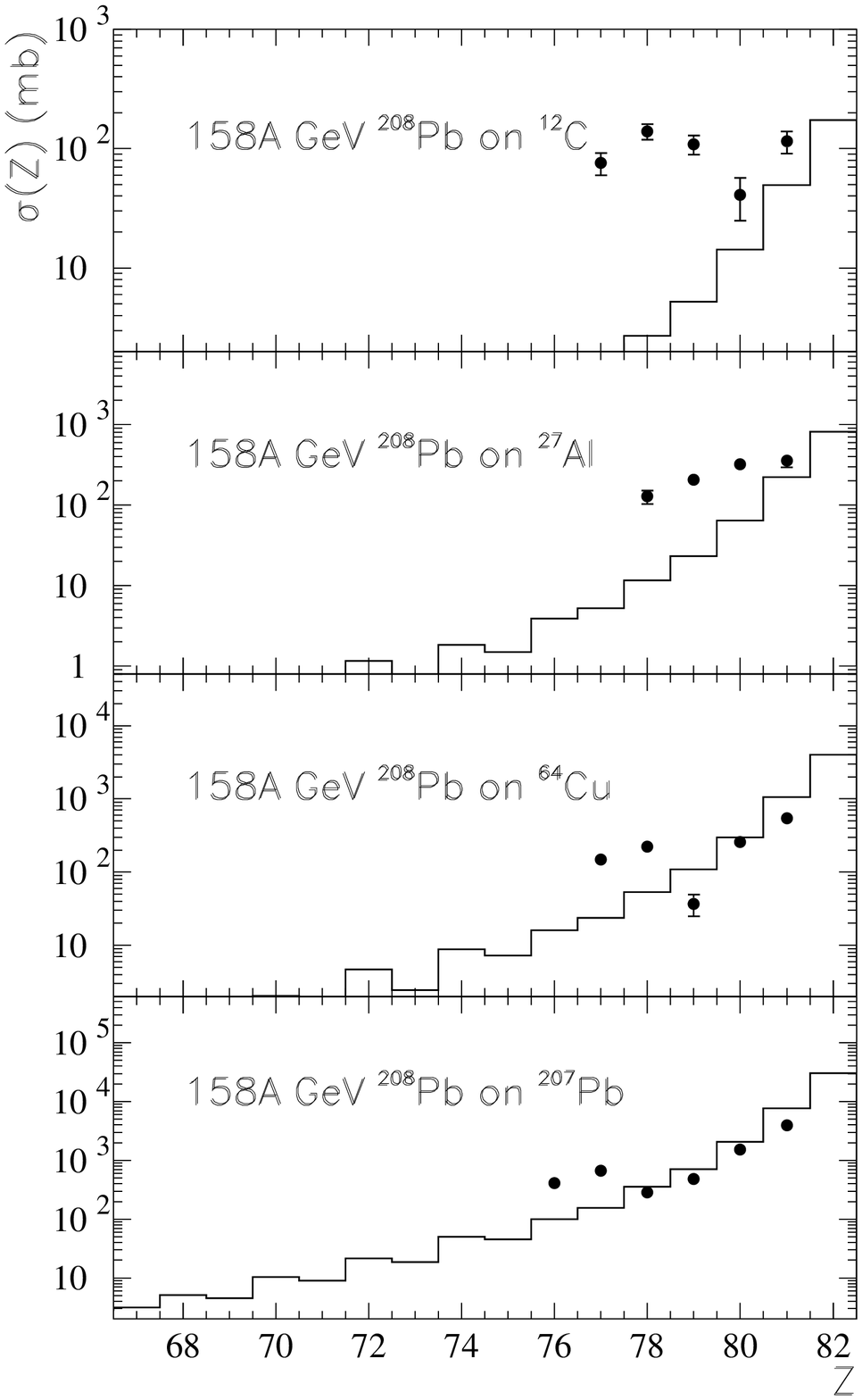}}
\caption{RELDIS results for fragmentation charge changing cross sections of 
$^{208}{\rm Pb}$ ions at SPS. Points: 
experimental data~\protect\cite{Dekhissi}.}
\label{fig:zdis}
\end{minipage}
\hspace{\fill}
\begin{minipage}[t]{65mm}
{\includegraphics[width=0.95\textwidth]{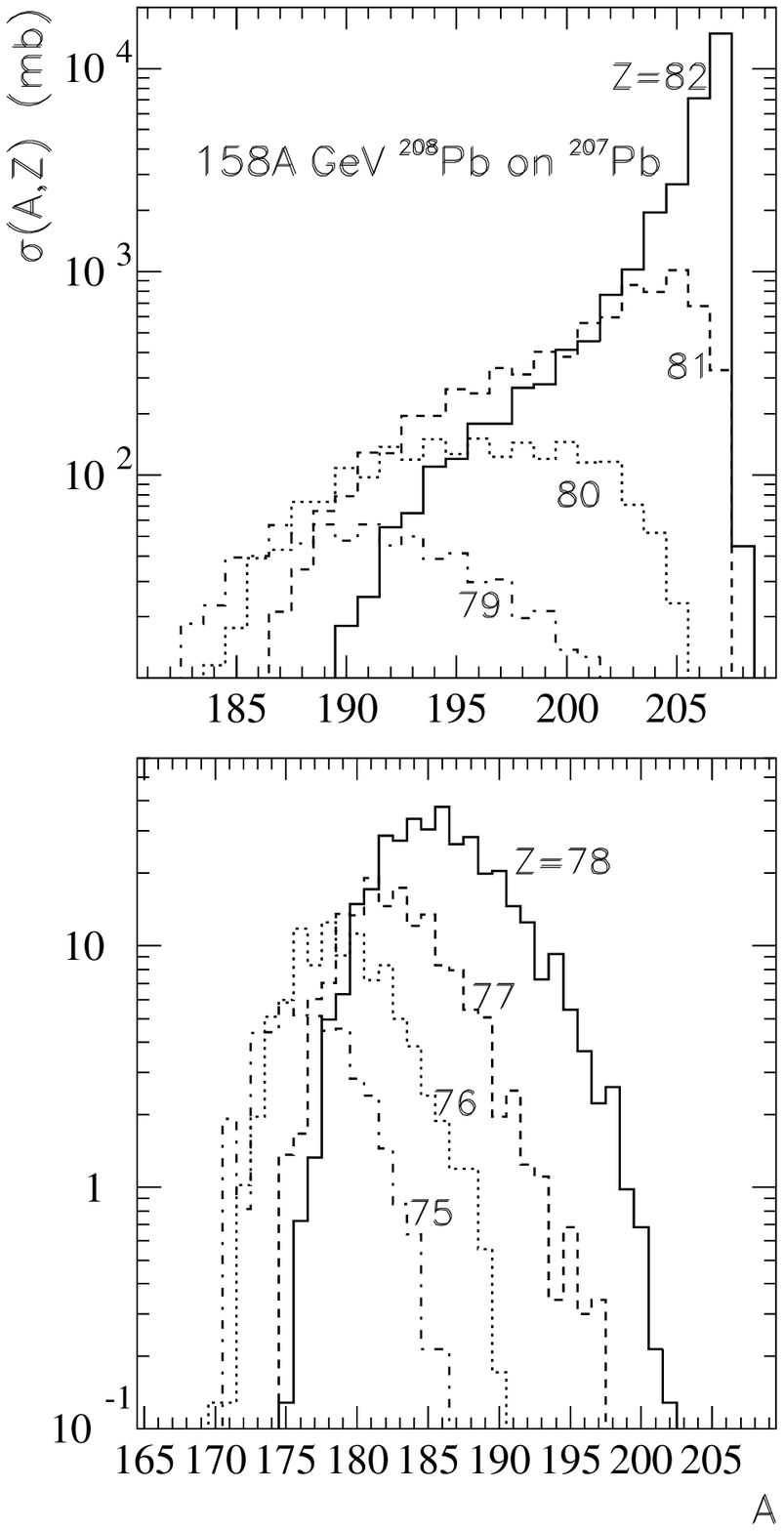}}
\caption{RELDIS predictions for mass distributions of isotopes 
with charges $Z=75,76,...,82$  produced in the 
electromagnetic dissociation of $^{208}{\rm Pb}$ ions at SPS.}
\label{fig:zdis_neut}
\end{minipage}
\end{figure}

Partial break-up charge changing cross sections ($-7\leq \Delta Z\leq 1$) 
for 158 AGeV $^{208}{\rm Pb}$ ions interacting with various targets, 
from H to Pb, were measured recently~\cite{Dekhissi}. 
Although the contributions to fragmentation due to the 
electromagnetic and nuclear forces were not separated in the 
experiment~\cite{Dekhissi}, one can use RELDIS code to evaluate the role of 
the electromagnetic dissociation for each particular dissociation mode.
As it is shown in Fig.~\ref{fig:zdis}, the main part of charge changing cross 
sections in PbPb collisions may be explained by the electromagnetic 
interaction.

Calculations show that multiple neutron emission 
also plays a role in each of the partial charge changing channels 
with $\Delta Z=-6,...,-1$ for PbPb collisions, see Fig.~\ref{fig:zdis_neut}. 
The lost of each proton is accompanied with a high 
probability by the multiple neutron emission.
For example, up to 20 neutrons may be emitted in  $\Delta Z=-2$ fragmentation 
channel, and the emission of 8--10 neutrons is the most probable process.   

This is evident from the fact
that due to a high Coulomb barrier of Pb nucleus, photoemission of  
protons takes place well above the GDR region, particularly
in the region of quasideuteron absorption, where the excitation energy, 
$E^\star\sim 20-50$ MeV, is sufficient to evaporate many neutrons.
On the contrary, the mass distribution for $\Delta Z=0$ channel is completely 
different since in this case $1n$ and $2n$ emission in GDR region strongly 
dominates. However, a long tail of the multiple neutron emission also
exists in this channel.

Our expectation of the intense neutron emission in the charge changing 
electromagnetic collisions of heavy nuclei is very different from a simple
picture assumed in Ref.~\cite{Hirzebruch}, where 
the rates of neutron and proton emission above the photoproton threshold
were determined by $N/Z$ ratio of the nucleus which undergoes fragmentation.
A detailed understanding of the fragmentation mechanism may be
obtained by measurements of fragment masses and detection of neutrons 
produced in collisions of ultrarelativistic heavy ions.

\section{Conclusions}

A variety of fragmentation mechanisms takes place in
the electromagnetic dissociation of ultrarelativistic heavy ions, namely 
the coalescence, Fermi break-up, evaporation, fission and multifragmentation. 
A partial or complete disintegration of the colliding nuclei 
is possible despite the absence of geometrical overlap of the nuclear 
densities.   
Comparison with CERN SPS data on PbPb collisions demonstrates the dominance of 
electromagnetic dissociation in the partial charge-changing fragmentation 
channels with $\Delta Z=-6,...,-1$. RELDIS model predicts also a very 
intense neutron emission in such channels.  

\ \\
\indent The author gratefully acknowledges the fruitful collaboration with 
A.S.Bot\-vina, J.P.Bondorf, A.S.Iljinov and I.N.Mishustin on the subjects
of this talk. Discussions with A.B.Kurepin, G.Giacomelli, M.Giorgini, 
L.Patrizii and P.Serra are greatly appreciated.
The author is very indebted to the Organizing Committee 
of the KEK-Tanashi Symposium for the kind hospitality and financial support.

\end{document}